\documentstyle[aps,pre,preprint]{revtex}
\begin{document}
\draft
\title{A non trivial extension of the two-dimensional Ising model:\\
the $d$-dimensional ``molecular'' model.}
\author{Fabio Siringo}
\address{Dipartimento di Fisica e Astronomia, ed 
Unit\`a INFM dell' Universit\`a di Catania,\\
Corso Italia 57, I 95129 Catania, Italy}
\date{\today}
\maketitle
\begin{abstract}
A recently proposed molecular model is discussed as a non-trivial extension
of the Ising model. For $d=2$ the two models are shown to be equivalent,
while for $d>2$ the molecular model describes a peculiar second order
transition from an isotropic high temperature phase to a low-dimensional
anisotropic low temperature state. The general mean field analysis is
compared with the results achieved by a variational Migdal-Kadanoff
real space renormalization group method and by standard Monte Carlo
sampling for $d=3$. By finite size scaling the critical exponent has
been found to be
$\nu=0.44\pm 0.02$ thus establishing that the molecular model
does not belong to the
universality class of the Ising model for $d>2$.
\end{abstract}


\section{Introduction}
The molecular model has been first proposed \cite{letter}
as a very simple $d$-dimensional
lattice model which incorporates some degree of frustration and thus
describes some aspects of molecular orientation in covalently bound
molecular solids.
Most molecular liquids retain their molecular structure even in
the solid phase, where some long range order usually shows up as
a consequence of inter-molecular interaction. However in the solid
the orientational order of the molecules may change according to
the thermodynamic conditions giving rise to quite rich phase diagrams
as recently observed for hydrogen under high pressure\cite{hydrogen}.
Moreover orientational
ordering is responsible for several phase transitions occurring even
in the liquid phase (liquid crystals), and lattice models have been found
to be useful for describing such transitions.

The most studied models describe a molecular interaction which
arises from dipole fluctuation,
is weak and gives rise to the observed
three-dimensional ordering of most molecular Van der Waals solids.
The thermodynamic behaviour of such weakly interacting systems
can be analyzed in terms of $O(3)$ symmetric vectorial models.

Conversely the molecular model was motivated by a description of the
almost-covalent molecular solids where the interaction has a covalent
main component and is characterized by some level of frustration
(since the coordination number for the covalent bond is quite low).
In such solids
each molecule must choose a few partners and cannot accept any
further invitation. The lower is the allowed coordination number,
the higher the frustration which gives rise to the low-dimensional
structures observed in polymers (one-dimensional) or in
iodine\cite{iodine} and hydrogen halides\cite{obriot,holzapfel}
(two-dimensional).
In iodine, where the covalent nature
of the interaction is out of doubt\cite{iodine1,iodine2}, even
one-dimensional zigzag
chains have been reported under pressure\cite{Pasternak}.
Moreover we expect that a
covalent interaction should show up for all the molecular solids
under high pressure, as the inter-molecular distance approaches the
intra-molecular length, provided that some important structural
transition does not occur first (like dissociation).
Quite recently the existence of charge transfer between $O_2$ molecules
has also been reported under pressure\cite{gorelli}

The molecular model is 
a simple frustrated lattice model which can
describe some aspects of a covalently bound molecular
solid. It consists of a $d$-dimensional hypercubic lattice with a
randomly oriented linear molecule at each site. In its simplest version
each molecule is only allowed to be oriented towards one of its
nearest neighbours. There is an energy gain for any pair of neighbours
which are oriented along their common bond (a covalent bond).
The existence of preferred orientational axes breaks the rotational
invariance of the single molecule as it is likely to occur for any
real molecular system under pressure.
In fact even in
hydrogen, the broken symmetry phase transition which
is observed under pressure\cite{silvera} has been recently
shown\cite{hemley} to be affected by the presence
of a crystal field which breaks the isotropy.

Quite recently, similar lattice models have been used for describing the
diffusion of particles and molecules inside a polymer, and the growth
of one-dimensional islands (polymeric chains)\cite{polymer}.
The molecular model has already stimulated some recent work on molecular
orientation in nitrogen\cite{tozzini,march}
which goes back to the phenomenology laid down
by Pople and Karasz\cite{pople}. It had been argued\cite{vega}
that the weak
intermolecular bonds between two $N_2$ molecules should 
be favourable for the formation of an orientationally disordered
``plastic crystal'' solid phase, and should lead to freezing
into an orientationally ordered phase. More recent experimental data
on nitrogen\cite{bini} confirm the existence of an orientational
disordering temperature in the solid below the melting temperature.
However, as far as we know, the molecular systems which are 
more closely described by the molecular model are the hydrogen halides
$HX(X=F,Cl,Br,I)$. Their low-temperature structures are known to
consist of planar chains of molecules in the condensed state while
a totally disordered structure has been observed with increasing
temperature at ambient pressure\cite{obriot}. Moreover the
opposite transition, from orientational disorder to an ordered chain
structure, 
has also been reported by increasing pressure\cite{holzapfel}.

The molecular model undergoes a transition from an
high-temperature (or weakly interacting) fully isotropic disordered
system, to a low-temperature (or strongly interacting) anisotropic
low dimensional broken-symmetry phase. As a consequence of frustration
the breaking of symmetry is accompanied by a sort of decomposition
of the system in low-dimensional almost independent parts, as observed
in solid iodine and hydrogen halides.
Such remarkable behaviour requires a space dimension
$d>2$, while for $d=2$ the model is shown to be equivalent 
to the exactly solvable two-dimensional Ising model\cite{Itz}. As shown
by Monte Carlo calculations, in the broken-symmetry phase the system 
displays the presence of correlated chains of molecules (polymers) which
point towards a common direction inside each two-dimensional sub-set of
the lattice (plane). Such planes are weakly correlated in the
low-temperature phase, and the system has a two-dimensional behaviour even
for $d>3$.

In this paper the relevance of the molecular model as a non-trivial
extension of the Ising model is pointed out. Thus, apart from the physical
motivations, the model is fully examined and the phase transition
is described by several methods: mean-field, real space
renormalization group and numerical simulations. Exactly solvable models are
important for our understanding of more complex systems, and provide a
test for approximate techniques. The $d$-dimensional molecular model
shares with the Ising model the $d=2$ realization, since their equivalence
for $d=2$ has been proven to be exact\cite{letter}.
In this paper we will focus on the
$d=3$ model, but we will take advantage of the existence of an exactly
solvable realization for $d=2$. For $d>2$, as the frustration increases,
the model shows a very different behaviour compared to the Ising or
Potts\cite{Potts} models. These last show a fully $d$-dimensional
broken-symmetry phase while the molecular model is characterized by a
low-dimensional ordering inside the planes with negligible
correlation among different planes. Moreover for
$d=3$ the
molecular model is shown to belong to a different universality class, since
its critical exponent $\nu$ turns out to be $\nu=0.44\pm0.02$
by finite size scaling.  We expect that such new universality class
should describe a broad group of isotropic physical systems characterized
by a low-dimensional ordering in their low-temperature phase.
Such broad class of phase transitions
should be explored by experiments in
order to compare with the theoretical predictions for the critical
universal properties. In such respect the driving parameter does not need
to be the temperature, as the bond strength can be directly modified by a
change of pressure in several systems.

This paper is organized as follows. Section II contains a formal definition
of the $d$-dimensional molecular model, and a proof of its
equivalence to the Ising model for $d=2$. In Section III the mean-field
solution is discussed for the generic $d$-dimensional model. In Section IV
a modified variational Migdal-Kadanoff method is presented and its
application is discussed for $d=2$ and $d=3$. At variance with a previous
calculation\cite{MK} which yielded a quite poor result, the variational
method is shown to work very well provided that some assumptions are
made on the nature of the broken phase.
Section V contains the
results of a numerical simulation by Monte Carlo sampling, and the
numerical estimation of both critical temperature and exponent by
finite size scaling. In Section VI the main findings are summarized
and discussed.

\section{The molecular model}

The molecular model was first described in Ref.\cite{letter}.
We briefly describe its formal definition in order to fix the notation.
Let us consider a $d$-dimensional hypercubic lattice, with a
randomly oriented linear molecule at each site. The molecules are
supposed to be symmetric with respect to their centre of mass
which is fixed at the lattice site. 
Only a discrete number of space orientations are allowed for
each molecule: we assume that each of them must point towards
one of its $2d$ first neighbour sites. This choice can be justified
by the existence of covalent interactions along preferred axes.
Then each molecule has $d$ different
states corresponding to molecular orientation along the hypercube
axes (molecules are symmetric). Finally, each couple
of first neighbour molecules, when pointing the one towards the other,
are assumed to gain a bonding energy for their directional covalent
bond (they touch each other). As shown in Fig.1 for $d=2$, bonding
in a direction excludes any possible bond along the other $(d-1)$
directions. The coordination number is $2$ for any value of $d$, and
the frustration increases with increasing $d$.

According to such description we introduce a versor variable
$\hat w_{\bf r}$ for each of the $N$ sites ${\bf r}$ of the lattice,
with $\hat w_{\bf r} \in\{\hat x_1, \hat x_2,\cdots \hat x_d\}$ 
pointing towards one of the $d$ hypercube axes $x_\alpha$.
The versors $\hat x_\alpha$ are assumed to be orthonormal:
$\hat x_\alpha \cdot \hat x_\gamma=\delta_{\alpha\gamma}$.
The partition function follows
\begin{equation}
Z=\sum_{\{\hat w\}} e^S=
\sum_{\{\hat w\}} \exp\left[{4\beta \sum_{{\bf r},\alpha} 
({\hat w_{\bf r}\cdot \hat x_\alpha})  
({\hat w_{{\bf r}+\hat x_\alpha}\cdot \hat x_\alpha})}\right]
\end{equation}
where $\{\hat w\}$ indicates a sum over all configurations,
$\alpha$ runs from $1$ to $d$, and the lattice spacing is set to
unity.
The inverse temperature $\beta$ (in units of binding energy)
can be negative for a {\it repulsive} model, but is assumed positive
in the molecular context.

The model may be generalized by introducing an external $d$-dimensional
vectorial field
${\bf h}(\alpha)$  at each link. The dependence on $\alpha$ means that
the field differs
according to the space direction $\alpha$ of the
lattice link which joins the sites. The modified partition function reads
\begin{equation}
Z_h=\sum_{\{\hat w\}} e^{S_h}=
\sum_{\{\hat w\}}
\exp\left\{ {4\beta \sum_{{\bf r},\alpha}\left[
\left(\hat w_{\bf r} \cdot \hat x_\alpha\right) 
\left(\hat w_{{\bf r}+\hat x_\alpha} \cdot \hat x_\alpha\right)
+ {\bf h}(\alpha)\cdot\hat w_{\bf r}
+ {\bf h}(\alpha)\cdot \hat w_{{\bf r}+\hat x_\alpha}\right]}
\right\}
\end{equation}

It is evident that if the field satisfies the condition
\begin{equation}
\sum_{\alpha} {\bf h}(\alpha)=0
\label{h=0}
\end{equation}
then $S_h$ does not depend on ${\bf h}$ and $S_h \equiv S$. In such
case the extra degree of freedom provided by ${\bf h}$ can be regarded
as a sort of internal symmetry of the model. This global symmetry can
be made local by allowing the field ${\bf h}$ to depend on the site
position ${\bf r}$. We will only take advantage of the global
symmetry in this paper. We notice that such symmetry cannot be seen
as a gauge invariance, since in lattice gauge models any gauge change
leaves the energy gain unchanged at any link. Here the field ${\bf h}$
changes the energy gain of all the links while the whole action is invariant.

Adopting a more compact notation, the partition function reads
\begin{equation}
Z_h=\sum_{\{\hat w\}} e^{S_h}=\sum_{\{\hat w\}} 
e^{\sum_{{\bf r},\alpha} {\cal L}({\bf r},\alpha)}
\end{equation}
where the Lagrangian density ${\cal L}$ follows as
\begin{equation}
{\cal L}({\bf r},\alpha)= \hat w_{\bf r}^\dagger M_\alpha(\beta, {\bf h})
\hat w_{{\bf r}+\hat x_\alpha}
\label{L}
\end{equation}
Here the canonical $d$-dimensional column vector representation  
of $R^d$ is employed with $\hat x_1 \equiv (1,0,0\dots)$,
$\hat x_2 \equiv (0,1,0\dots)$, etc. The $d\times d$ matrix $M_\alpha$
does not depend on the configurations of the system, and entirely
characterizes the model. 

The global symmetry of the action provides a simple way to show the
equivalence between molecular and Ising models for $d=2$.
For the two-dimensional lattice the condition (\ref{h=0}) is satisfied
by the field ${\bf h} (1)=h(\hat x_1
-\hat x_2)$, ${\bf h} (2)=-{\bf h}(1)$.
The matrix $M_\alpha$ follows
\begin{eqnarray}
M_1&=&\left(\matrix{4\beta(1+2h) & 0 \cr 
             0            & -8\beta h\cr}\right),
\nonumber\\
M_2&=&\left(\matrix{-8\beta h & 0 \cr 
                   0        & 4\beta (1+2h) \cr}\right).
\label{Md2}
\end{eqnarray}
Then for $h=-1/4$, $M_1\equiv M_2$, and ${\cal L}$ reads
\begin{equation}
L({\bf r}, \alpha)=\beta+\hat w_{\bf r}^\dagger
\left(\matrix{\beta & -\beta \cr -\beta & \beta\cr}\right)
\hat w_{{\bf r}+\hat x_\alpha}.
\end{equation}
Identifying the two-dimensional column versors $\hat w$ with spin variables,
apart from an inessential factor, $Z$ reduces to the partition
function of a two-dimensional Ising model
\begin{equation}
Z=e^{2\beta N}\cdot Z_{Ising}
\end{equation}
and is exactly solvable.
For $\beta \to +\infty$ a ground state is approached with all the
molecules oriented along the same direction, and with formation of
one-dimensional polymeric chains (Fig.2a); for $\beta \to
-\infty$ the repulsive model approaches a zero-energy (no bonds)
ground state analogous to the antiferromagnetic configuration of
the Ising model (Fig.2b).

For $d\ge 3$ the analogy with the Ising model breaks down, and this
is evident from a simple analysis of the ground state configuration.
Due to frustration the model has an infinitely degenerate ground state
in the thermodynamic limit $N\to \infty$. For instance, in the case
$d=3$, the minimum energy is obtained by orienting all the molecules
along a common direction, as for $d=2$. However the ground state
configuration is not unique: the number of molecular bonds does not
change if we rotate together all the molecules belonging to an entire
layer which is parallel to the original direction of orientation.
As a consequence of frustration the total degeneration is
$3(2^{(N^{1/3})})$, and the system could even behave like a glass
for the large energy barriers which separate each minimum from the
other. The phase diagram is expected to be quite rich, with at least
a transition point between the high temperature disordered phase
and an ordered broken-symmetry low temperature phase. 

\section{ Mean Field Approximation}

For the generic $d$-dimensional model, some analytical results can be
obtained in Mean-Field (MF) approximation:
neglecting second order fluctuation terms
\begin{equation}
(\hat w_{\bf r} \cdot \hat x_\alpha) 
(\hat w_{{\bf r}+\hat x_\alpha} \cdot \hat x_\alpha)\approx
\Delta_\alpha
(\hat w_{\bf r} \cdot \hat x_\alpha)+
\Delta_\alpha
(\hat w_{{\bf r}+\hat x_\alpha}\cdot \hat x_\alpha)
-\Delta_\alpha^2
\end{equation}
where $\Delta_\alpha=\langle \hat w_{\bf r}
\cdot\hat x_\alpha\rangle$ is an average over the configurations,
and $\sum_\alpha
\Delta_\alpha=1$ (with the obvious bounds $0\le\Delta_\alpha\le 1$).
Here the order parameter $\Delta_\alpha$ gives the probability of finding
a molecule oriented along the direction of $\hat x_\alpha$.
The partition function factorizes as
\begin{equation}
Z_{MF}=
\left(\sum_\alpha e^{8\beta\Delta_\alpha}\right)^N
\exp\left(-4N\beta\sum_\alpha \Delta_\alpha^2\right)
\end{equation}
and the free energy follows
\begin{equation}
F_{MF}=-{1\over{N\beta}} \log Z_{MF}=4\sum_\alpha \Delta_\alpha^2
-{1\over\beta}\log\left(\sum_\alpha e^{8\beta\Delta_\alpha}\right).
\label{fmf}
\end{equation}
The derivative with respect to $\Delta_\mu$ yields, for the stationary
points,
\begin{equation}
\Delta_\mu={{e^{8\beta\Delta_\mu}}\over{\sum_\alpha e^{8\beta\Delta
_\alpha}}}
\label{del}
\end{equation}
which satisfies the condition $\sum_\alpha\Delta_\alpha=1$.

In the high temperature limit $\beta\to 0$ eq.(\ref{del})
has the unique
solution $\Delta_\mu=1/d$ which reflects the complete random orientation
of molecules. In the opposite limit $\beta\to\infty$, apart from such
solution, eq.(\ref{del}) is satisfied by the broken-symmetry field 
$\Delta_\mu=1$, $\Delta_\alpha=0$ for $\alpha\not=\mu$, which obviously
corresponds to a minimum for $F_{MF}$. Then at a critical point 
$\beta=\beta_c$ the high temperature solution must become unstable
towards a multivalued minimum configuration.
The Hessian matrix is easily evaluated at the stationary points
by using eqs.(\ref{del}) and ({\ref{fmf}):
\begin{equation}
H_{\mu \nu}={1\over 8} {{\partial^2 F_{MF}}\over{\partial\Delta_\mu
\partial\Delta_\nu}}=\delta_{\mu \nu}\left(1-8\beta\Delta_\mu\right)
+8\beta\Delta_\mu\Delta_\nu
\end{equation}
In the high temperature phase ($\beta<\beta_c$),
inserting $\Delta_\mu=1/d$,
the eigenvalue problem
\begin{equation}
\det\vert H_{\mu\nu}-\lambda\delta_{\mu\nu}\vert=0
\end{equation}
yields
\begin{equation}
\left(1-{{8\beta}\over{d}}-\lambda\right)^{d-1}\cdot\left(1-\lambda
\right)=0.
\end{equation}
Thus the Hessian matrix is positive defined if and only if         
$\lambda=(1-8\beta/d)>0$. Beyond the critical point $\beta=\beta_c=
(d/8)$ the solution $\Delta_\mu=1/d$ is not a minimum, and a multivalued
minimum configuration shows up. Such result obviously agrees with the
MF prediction for the Ising model, $\beta_{Ising}=1/(2d)$, only for
the special dimension $d=2$. For $d>2$ we observe an increase of
$\beta_c$ with $d$, to be compared to the opposite trend shown by the
Ising model. Such behaviour may be interpreted in terms of the low
dimensionality of the ordered phase. Due to frustration the ordering
may only occur on a low dimensional scale: for instance in three
dimensions each layer has an independent internal ordering. Thus we
expect a larger $\beta_c$ for $d>2$ since the increasing of $d$ only
introduces larger fluctuations, with each molecule having $(d-2)$
allowed out-of-plane orientations. For $d=3$ the low temperature
phase can be regarded as a quenched disordered superposition of
layers which are internally ordered along different in-plane
directions. As a consequence of frustration the system shows
a two-dimensional character below the critical point while behaving as
truly three-dimensional in the high temperature domain.
In MF the neglecting of some fluctuations usually leads to a
critical temperature which overestimate the exact value (i.e. the
critical inverse temperature $\beta_c$ is underestimated).
For $d=2$
the MF prediction is $\beta_c=0.25$ to be compared with the exact
value $\beta_c=0.4407$. For $d=3$ the MF prediction 
$\beta_c=d/8=0.375$ should provide a lower bound to the unknown
exact value.

\section{Variational Migdal-Kadanoff Approximation}

According to the Migdal-Kadanoff\cite{Itz,Mig+Kad}
method, a link displacement may
be introduced by considering that the configurational average of the
Lagrangian density ${\cal L}$ in eq.(\ref{L})
must be translationally invariant
\begin{equation}
\langle {\cal L}({\bf r},\alpha)\rangle=
\langle {\cal L}({\bf r^\prime},\alpha)\rangle
\end{equation}
then defining
\begin{equation}
\Gamma_\alpha({\bf r}, {\bf r^\prime})={\cal L}({\bf r}, \alpha)-
{\cal L}({\bf r^\prime},\alpha)
\end{equation}
we can state that 
$\langle \Gamma_\alpha({\bf r}, {\bf r^\prime})\rangle=0$ and the same
holds for any sum $\Gamma$ over an arbitrary set of such terms
\begin{equation}
\Gamma=\sum \Gamma_\alpha ({\bf r}, {\bf r^\prime}).
\label{g}
\end{equation}
Replacing the action $S_h$ by the sum $S_h+\Gamma$, and assuming that the
condition (\ref{h=0}) is verified
(so that we can drop the $h$ in $S_h$ and
$Z_h$ which are invariant), the modified partition function $Z_\Gamma$
can be approximated by cumulant expansion as
\begin{equation}
Z_\Gamma=\sum_{\{\hat w\}} e^{S+\Gamma}=Z\cdot\langle e^\Gamma\rangle
\approx Z\left[ e^{\langle\Gamma\rangle}\cdot e^{{1\over 2}
\left(\langle \Gamma^2\rangle-\langle\Gamma\rangle^2\right)
}\right]
\end{equation}
then, since $\langle \Gamma\rangle=0$,
\begin{equation}
Z_\Gamma\approx Z\cdot e^{{1\over 2}\langle\Gamma^2\rangle}
\label{zg}
\end{equation}
For instance, the sum in equation (\ref{g})
could run over all $\alpha\not=1$,
and for appropriate values of the vectors ${\bf r}, {\bf r^\prime}$, in
order to yield a displacement of links which are orthogonal to $\hat x_1$.
To second order in $\Gamma$, the error introduced by link displacement
is controlled by the exponential factor in equation (\ref{zg}).

Link displacement breaks the internal symmetry of the model, so that
$Z_\Gamma$ is no longer invariant for any field change subject to the
condition (\ref{h=0}).
Then we may improve the approximation by using the extra
freedom on the choice of ${\bf h}$ for minimizing the difference between
the approximate partition function $Z_\Gamma$ and the exact $Z$.

If ${\bf h}$ satisfies the condition (\ref{h=0})
then $a{\bf h}$ satisfies such
condition as well for any choice of the scalar parameter $a$.
Then a special class of invariance transformations can be described by
a change of the strength parameter $h$,
assuming the field ${\bf h}$ as proportional
to $h$. The following
discussion could be easily generalized to other classes of transformations
described by more than one parameter.
Since $\Gamma$ is linear in the field ${\bf h}$, then in general
\begin{equation}
\Gamma^2=\left[A+hB\right]^2
\end{equation}
where $A$ and $B$ depend on the configuration of the system.
For the average we have
\begin{equation}
\langle \Gamma^2\rangle=\langle A^2\rangle+2h \langle AB\rangle +
h^2 \langle B^2\rangle
\label{g2a}
\end{equation}
This last equation, inserted in eq.(\ref{zg}) leads to the
following considerations:
i) the coefficient $\langle B^2\rangle$
is positive defined, then the average $\langle \Gamma^2\rangle$ always
has a minimum for an appropriate value of $h=h_0$; ii) in general
$\langle AB\rangle \not=0$ then $h_0\not=0$, and a direct use of the
Migdal-Kadanoff method on the original model (with no field considered)
would yield a larger error; iii) to the considered order of approximation
$Z_\Gamma$ is stationary at $h=h_0$, and is symmetric around that point,
then all the physical properties described by such partition function
must result symmetric with respect to $h_0$. Moreover, at the same order
of approximation, any physical
observable $f$ will acquire an unphysical dependence on $h$, and the
symmetry around $h_0$ requires that ${{df}\over{dh}}=0$ for $h=h_0$.
Then we expect that all such observables
should be stationary at $h=h_0$, and their best estimate should coincide
with the extreme value.

As a consequence of the above statements, the Migdal-Kadanoff method
can be improved by taking advantage of the global symmetry of the
model. By use of the approximate
partition function $Z_\Gamma$ the critical temperature acquires
a non-physical field dependence, but the best estimate of $\beta_c$ is its
stationary value corresponding to $h=h_0$. The method can be seen as a
variational method with the best approximation achieved by the minimum
in the inverse temperature.

Such stationary condition resembles the principle of
``minimum sensitivity'' introduced by Stevenson\cite{Stevenson} for
determining the best renormalization parameters whenever the physical
amplitudes depend on them (and they should not). In our context, since
the critical temperature should not depend on the choice of the field
strength $h$, the best value for such field is the one which makes the
critical temperature less sensitive i.e. the stationary point.
However, according to equation (\ref{zg}) and
(\ref{g2a}), here we have got 
a formal proof of the stationary condition up to second order of the
cumulant expansion.

The method may be used by performing a
displacement of links that are orthogonal to $\hat x_1$, and then 
a one-dimensional decimation along the $\alpha=1$ axis.
According to such program
let us define the alternative $d\times d$ matrix
$t_\alpha (\beta, h)$ 
\begin{equation}
e^{{\cal L}({\bf r},\alpha)}=
\hat w_{\bf r}^\dagger t_\alpha(\beta, h)
\hat w_{{\bf r}+\hat x_\alpha}
\end{equation}
The partition function follows 
\begin{equation}
Z_h=\sum_{\{\hat w\}} \prod_{{\bf r},\alpha}
\left[\hat w_{\bf r}^\dagger t_\alpha(\beta, h)
\hat w_{{\bf r}+\hat x_\alpha} \right]
\end{equation}
After link displacement and decimation along the $\alpha=1$
axis, the modified partition function
reads
\begin{equation}
Z_\Gamma=\sum_{\{\hat w\}} \prod_{{\bf r},\alpha}
\left[\hat w_{\bf r}^\dagger\tilde t_\alpha(\beta, h)
\hat w_{{\bf r}+\hat x_\alpha} \right]
\end{equation}
where the sum and the product run over the configurations and the 
sites of the new decimated lattice, and
\begin{equation}
\tilde t_1(\beta, h)=\left[ t_1(\beta, h)\right]^\lambda
\end{equation}
\begin{equation}
\tilde t_\alpha (\beta, h)=t_\alpha (\lambda\beta, h) \qquad
{\rm for}\quad \alpha \not=1
\end{equation}
with $\lambda$ being the scale factor between the new and the old lattice.
A renormalized inverse temperature $\tilde \beta_\alpha$ 
may be defined according
to
\begin{equation}
\tilde t_1(\beta, h)=t_1 (\tilde \beta_1, h)
\label{bet1}
\end{equation}
\begin{equation}
\tilde \beta_\alpha=\lambda\beta\qquad {\rm for}\quad\alpha\not=1
\label{bet2}
\end{equation}
Eventually, the same scaling operation should be performed consecutively
for all the directions in order to obtain an hyper-cubic lattice again. 
For any finite scaling parameter $\lambda>1$ the renormalized inverse
temperature is anisotropic, but an isotropic fixed-point can be recovered
in the limit $\lambda\to 1$. The equations (\ref{bet1}),(\ref{bet2})
define the flow of the
renormalized inverse temperature, which changes for any different value
of the field strength $h$. Equation (\ref{bet1})
has a more explicit aspect in
the representation of the common eigenvectors of the matrices $t_1$ and 
$\tilde t_1=[t_1]^\lambda$. The rank of such matrices is 2 for any space
dimension $d$, as can be expected from the definition of the model. Then
both the matrices can be represented in terms of the two non-vanishing
eigenvalues $\eta_1$, $\eta_2$, which are functions of $\beta$ and $h$.
Assuming that $\eta_2\not=0$, and defining 
\begin{equation}
f(\beta,h)={{\eta_1}\over{\eta_2}}
\end{equation}
apart from a regular multiplicative factor for the partition function,
the scaling equation (\ref{bet1}) reads
\begin{equation}
\left[f(\beta,h)\right]^\lambda=f(\tilde\beta_1, h)
\end{equation}
For any fixed $h$, the fixed points follow through the standard
Migdal-Kadanoff equations
\begin{equation}
\left[f(\lambda^{\alpha-1}\beta_\alpha, h)\right]^\lambda=
f(\lambda^{\alpha-d}\beta_\alpha, h).
\label{MK}
\end{equation}
When $\lambda$ is analytically continued up to $1$ such equations
give the same isotropic fixed point $\beta_c$. In fact, the
expansion of equations (\ref{MK})
around $\lambda=1$ implies (up to first
order in $\lambda-1$)
\begin{equation}
\ln f(\beta_c, h)=-(d-1)\beta_c\left[
{1\over f}
{{df}\over{d\beta}} \right]_{\beta_c}
\label{fix}
\end{equation}
which is an implicit equation for $\beta_c$.
Such equations yield their best estimate of $\beta_c$ when the strength
of the field $h$ is set to the stationary value $h_0$.

It is instructive to evaluate the stationary point $h_0$ for the
case $d=2$ which is equivalent to the two-dimensional Ising model
for the choice $h=h_I=-1/4$, as shown in section II.
The $h$ invariance of the exact partition function
guarantees the equivalence
of the two models for any choice of $h\not=h_I$. However, the
mere application of the Migdal-Kadanoff equations (\ref{MK})
to the simple
$h=0$ molecular model fails to predict even the existence of
the fixed point. On the other hand, for $h=h_I$, the very same
recurrence equations (\ref{MK}) are known to predict the exact fixed
point in the limit $\lambda\to 1$. That can also be checked by
inserting in equation (\ref{fix})
the exact expression for the fixed point of
the two-dimensional Ising model.
Such contradictory results are not surprising since, as already discussed,
link displacement breaks the $h$ invariance of
the model, and the approximate solution is thus dependent on the choice
of $h$.
We would like to test the variational method on this exactly solvable
model: we look for the stationary point of the
function $f(\beta, h)$. The matrices $t_\alpha$ follow from
the equations (\ref{Md2})
\begin{eqnarray}
t_1&=&\left(\matrix{ x^2b     & 1 \cr
             1            & {x^{-2}} \cr}\right)
\nonumber\\
t_2&=&\left(\matrix{  {x^{-2}}     & 1 \cr 
                   1      & xb \cr}\right)
\end{eqnarray}
where $b=exp(4\beta)$, and $x=exp(4\beta h)$. Then for the eigenvalues
we obtain
\begin{equation}
f(\beta, h)={{\eta_1}\over{\eta_2}}={ {(bx^4+1)-\sqrt{(bx^4-1)^2+4x^4}}
\over{(bx^4+1)+\sqrt{(bx^4-1)^2+4x^4}} }
\end{equation}
It can be easily shown that if the derivative of $f$ is zero at a given
$h$ independent of $\beta$, then the solution $\beta_c$ of Eq.(\ref{fix})
is stationary at that $h$ value.
Differentiating with respect to $x$, we find that the derivative of $f$
vanishes for $x^4=1/b$, which yields  $h=-1/4=h_I$ for any $\beta$.
As expected, this is the required value in order to recover the Ising model.
Thus the Migdal-Kadanoff approximation gives an improving estimate of
the critical point as we move from the {\it molecular} towards the
{\it Ising} representation (where the approximation yields
the exact fixed point). We stress that all such representations are
equivalent due to the $h$ invariance of the action.

For $d>2$ 
no equivalence to standard
studied models has been found, and the behaviour seems to be
dictated by the strong frustration which does not allow an higher
coordination number than two, even for higher dimensions.
We will focus on the $d=3$ model in order to compare the results with
the Monte Carlo findings of the next section. First of all the fields
${\bf h}(\alpha)$ must be defined. An isotropic choice would be
\begin{eqnarray}
{\bf h}(1)&=&h(\hat x_1-{1\over 2}\hat x_2-{1\over 2}\hat x_3)
\nonumber\\
{\bf h}(2)&=&h(\hat x_2-{1\over 2}\hat x_3-{1\over 2}\hat x_1)
\nonumber\\
{\bf h}(3)&=&h(\hat x_3-{1\over 2}\hat x_1-{1\over 2}\hat x_2)
\end{eqnarray}
The matrix $t_1$ follows
\begin{equation}
t_1=\left(\matrix{e^{4\beta+8\beta h} & e^{2\beta h} & e^{2\beta h} \cr
             e^{2\beta h} & e^{-4\beta h} & e^{-4\beta h} \cr
             e^{2\beta h} & e^{-4\beta h} & e^{-4\beta h} \cr}\right)=
\left(\matrix{  bx^2  & \sqrt{x} & \sqrt{x} \cr 
           \sqrt{x}   & 1/x   &  1/x    \cr
           \sqrt{x}   & 1/x   &  1/x    \cr}\right)
\end{equation}
Then from the eigenvalues
we obtain
\begin{equation}
f(\beta, h)={{\eta_1}\over{\eta_2}}={ {(bx^3+2)-\sqrt{(bx^3-2)^2+8x^3}}
\over{(bx^3+2)+\sqrt{(bx^3-2)^2+8x^3}} }
\label{f}
\end{equation}
Differentiating with respect to $x$ we find that the derivative of $f$
only vanishes for $x^3=2/b$. This is equivalent to say 
\begin{equation}
h=h_m=-{1\over 3}+{1\over{12\beta}}\ln 2
\end{equation}
which depends on $\beta$.
For such field strength $h_m(\beta)$ the ratio between the eigenvalues
reduces to
\begin{equation}
f(\beta, h_m)=\tanh(\beta)
\label{fi}
\end{equation}
which is exactly the same expression holding
for the Ising model\cite{Itz}.
However we must point out that in such case $h_m$ is not the stationary
point $h_0$. Since $h_m$ depends on $\beta$, the vanishing of the
derivative of $f$ does not imply that the solution of eq.(\ref{fix}) is
stationary. In fact, for $d=3$, the choice $h=h_m$ yields the known
poor result $\beta_c=0.1398$ by insertion of Eq.(\ref{fi})
in Eq.(\ref{fix}). On the other hand, by insertion of the general expression
for $f$ Eq.(\ref{f}), the scaling equation (\ref{fix}) can be numerically
solved for $\beta_c$ as a function of $h$. At the stationary point
$\beta_c$ has a minimum, and thus the variational method yields an even
worse prediction ($\beta_c\approx 0.12$ at the stationary point).
These shortcomings show that the isotropic $d=3$ variational method does
not suite the molecular model. Actually both MF and Monte Carlo methods
predict a larger $\beta_c$ and, as pointed out at the end of the
previous section,
the exact $\beta_c$ should be larger than the MF prediction
$\beta_{MF}=0.375$.

We could have guessed such disagreement since we are using an
isotropic version of the variational Migdal-Kadanoff method for a
system which is not isotropic in its ordered phase.
At the transition
point the system choices a direction, as is
usual for any symmetry breaking mechanism. However, at variance with
usual models, in the ordered phase the correlation length cannot be
isotropic: order occurs inside all layers which are orthogonal to the
chosen direction, while there is a negligible correlation along
such direction. It would be more sensible to
describe the ordering which takes place inside a single layer, thus
neglecting any correlation among different layers. Inside each layer
the correlation length is isotropic, and the $d=2$ variational
Migdal-Kadanoff method should give a better description of the transition.
The same argument should hold for the generic $d$-dimensional
molecular model. Moreover, despite the cost of this further approximation,  
the Migdal-Kadanoff method is known to work better for the lower dimensions,
and a $d=2$ variational method could provide a tool for
describing the generic $d$-dimensional molecular model even for $d>3$.

A $d=2$ version of the variational method requires a different choice for
the fields ${\bf h}(\alpha)$ which do not need to be isotropic any more.
Let us take the same field we used in section II, namely
${\bf h}(1)=h(\hat x_1-\hat x_2)$, ${\bf h}(2)=-{\bf h}(1)$ and
${\bf h}(3)=0$.
The matrix $t_1$ follows
\begin{equation}
t_1=\left(\matrix{e^{4\beta+8\beta h} & 1 & e^{4\beta h} \cr
                       1  & e^{-8\beta h} & e^{-4\beta h} \cr
             e^{4\beta h} & e^{-4\beta h} & 1 \cr}\right)=
\left(\matrix{  bx^2  & 1     &   x  \cr 
                 1    & 1/x^2 &  1/x \cr
                 x    & 1/x   &  1   \cr}\right)
\end{equation}
Notice that this is a $3\times 3$ matrix since we are using the $d=2$
method but we are still dealing with a $d=3$ molecular model.
The two matrices $t_1$ and $t_2$ share the same eigenvalues. Their
ratio is
\begin{equation}
f(\beta, h)={ {(bx^2+1+1/x^2)-\sqrt{(bx^2-1-1/x^2)^2+4(1+x^2)}}
\over{(bx^2+1+1/x^2)+\sqrt{(bx^2-1-1/x^2)^2+4(1+x^2)}} }
\end{equation}
Inserting this result in the scaling equation (\ref{fix}) evaluated
at $d=2$ yields
an implicit equation for $\beta_c$ versus $h$. The numerical solutions
are reported in Fig.3. They share most of the features of the $d=2$
molecular model: (i) There are several solutions but
there is no repulsive fixed point for $h=0$; (ii) The physical solution
starts at a negative $h$ which in this case is $h\approx -0.226$;
(iii) The physical solution has just one stationary point $h_0$ where
$\beta_c$ reaches its minimum value. However in this case the stationary
point is at $h_0=-0.2349$ where $\beta_c=0.6122$. This best estimate of
the critical point is not too far from the finite size scaling prediction
of the next section $\beta_c=0.53$. The result is encouraging, and gives
us more confidence in our understanding of the physics described by the
molecular model. Strictly speaking, this $d=2$ variational method
describes the transition occurring in a single layer of molecules.
However, at variance with the $d=2$ molecular model, each molecule is
now allowed to be oriented along three different axes (two in-plane
and one out-of-plane orientations). Thus this reasonable prediction
for $\beta_c$ could be regarded as an indirect proof that the
correlation between two different layers is negligible, and that in
the ordered phase the system behaves as a truly two-dimensional one.

\section{Monte Carlo sampling}

In order to check the prediction achieved by different approximate
methods it would be desirable to have an accurate numerical estimate
of the critical temperature. That can be easily obtained by finite size
scaling. Moreover, according to the scaling hypothesis, the critical
exponent $\nu$ can be extracted by the numerical data with a good
accuracy.

Cubic samples $N\times N\times N$ with $N=10,15,20,25,30$ have been
considered. All the averages have been evaluated by Monte Carlo sampling
with no special boundary conditions.

In this model
any ordering is characterized by the presence of some degree of
correlation along  one-dimensional chains of molecules. For $d=3$
there are $3\times N \times N$ different chains in each sample. Each
chain may be labelled by its direction $\alpha=1,2,3$ and by a couple
of integer
coordinates $I_1, I_2$  running over a lattice
layer orthogonal to the axis
$\hat x_\alpha$. For any chain we define an order parameter
\begin{equation}
m(\alpha,I_1,I_2)={1\over N}\sum_{J_\alpha=1}^N
\left[ \hat w(J_\alpha,I_1,I_2)\cdot \hat x_\alpha\right]
\end{equation}
where $\hat w(J_\alpha, I_1, I_2)$ is the versor $\hat w_{\bf r}$
at the chain site ${\bf r}$ whose integer coordinates are determined
by $J_\alpha$ along the chain and by the couple $I_1,I_2$ in the
orthogonal directions. If there is no correlation at all ($\beta\to 0$)
then $m(\alpha, I_1,I_2)\approx 1/3$ for any chain in the sample.
By averaging over all the chains of each sample and over all the
configurations, we obtain $\langle m\rangle=1/3$. For large $N$,
according to the central limit theorem, in this statistical ensemble the
variable $m$ follows a gaussian distribution centered at its average
value. In the opposite limit ($\beta\to\infty$) a third of the chains in
each sample have a large $m\approx 1$, while $m\approx 0$ for two thirds
of them. Since any intermediate value of $m$ is unlikely, the statistical
distribution of $m$ can be regarded as the superposition of two
different peaked distributions centered at $m=0$ and $m=1$.
If $N$ is large enough, and for a large number of configurations, such
distributions are very peaked and their width is very small.
Actually, just
below the critical point the gaussian distribution already
splits in a double-peak
distribution. We can monitor the transition by use of the new variable
$\gamma$
\begin{equation}
\gamma={{m-\langle m\rangle}
\over{\sqrt{\langle m^2\rangle-\langle m\rangle^2}}}
\end{equation}
By its definition the configurational average of $\gamma$ is vanishing
$\langle \gamma\rangle=0$ and the second moment
$\langle \gamma^2\rangle=1$.
The variable $\gamma$ only differs from $m$ for a shift and a rescaling,
thus the statistical distribution for $\gamma$ follows the same
trend already
discussed for $m$. However the fourth moment $s=\langle \gamma^4\rangle$ is
now strongly dependent on the number of peaks characterizing the
statistical distribution. For a single gaussian $s=3$
exactly.
Below the critical temperature the distribution splits. In the
thermodynamic limit $N\to\infty$ the width of each peak vanishes, while
the two peaks separate by a finite quantity. For instance assume that
just below the critical point a third of the chains yield
$m\approx 1/3+\epsilon$ where $\epsilon$ is a very small increase in the
chain correlation  which breaks the symmetry of the sample. The other
two thirds of chains must yield $m\approx 1/3-\epsilon/2$
since by its definition $\langle m\rangle=1/3$ exactly. Neglecting
the width of the peaks we may approximate the statistical distribution
for $m$ as the superposition of two delta-functions with weight factors:
\begin{equation}
P(m)={2\over 3}\delta (m-{1\over 3}+{\epsilon\over 2})
+{1\over 3}\delta (m-{1\over 3}-\epsilon)
\end{equation}
By use of such approximate statistical distribution
the calculation of the fourth moment $s=\langle \gamma^4\rangle$ is
straightforward and gives $s=1.5$ for any $\epsilon$,
no matter how small. This is one half of the single gaussian value.
Thus in the thermodynamic limit we expect that the fourth moment 
$s$ should behave like a step function with constant values
$s=3$ and $s=1.5$ respectively
above and below the critical temperature, and a sharp jump at the critical
point. For finite-size samples the fourth moment is expected to be
continuous across the transition, but according to the scaling
hypothesis the critical value should not depend on the sample size if we
assume a one-parameter scaling law across the critical point:
\begin{equation}
s=s\left(L/\xi(\beta)\right)
\label{sc}
\end{equation}
where $L$ is here the sample length, and $\xi$ is the correlation
length which is a function of temperature. According to such scaling
law $s=s(0)$ at the critical point for any $L$.

We have checked this
prediction by standard Monte Carlo sampling. For any fixed sample size,
we have taken a completely random initial configuration, and thermalized it
at a very high temperature ($\beta\approx 0.02$)
by $5\cdot 10^4$  complete sweeps. The temperature is then
decreased by steps of $\Delta \beta=0.02$. At each step a good
thermalization is achieved by $8\cdot 10^3$ complete sweeps, and then the
averages are evaluated over the successive $2\cdot 10^3$ sweeps.
Once a sufficiently low temperature is reached ($\beta\approx 1$), the
process is reversed and the temperature increased up to the initial value.
We have checked that the hysteresis is small in all the considered range
of temperature. Moreover the small differences observed going up and down
give a measure of the errors on the configurational
averages which have been approximated
by the mean values. The fourth moment $s$ is reported in Fig.4 for
$N=15,20,25,30$. All the curves cross at the same point
$\beta_c=0.53\pm 0.01$ as predicted by the one-parameter scaling
hypothesis. Moreover for very large or very small temperatures the
correlation length becomes very small and the fourth moment $s$
should approach its thermodynamic-limit value $s\to s(\infty)$ which is
expected to be $s(\infty)=3$  at high temperature and $s(\infty)=1.5$
at low temperature. As shown in Fig.4 the measured $s$ approaches
such limits far away from the critical point.

According to the usual definition of critical exponent
\begin{equation}
\xi\sim{1\over{(\beta-\beta_c)^\nu}}
\end{equation}
the scaling equation (\ref{sc}) allows for an accurate estimate of its
value:
linearizing $s$ around the critical point yields
\begin{equation}
\ln L=\nu\ln\vert s^\prime (\beta_c)\vert + {\rm const.}
\label{fit}
\end{equation}
where $s^\prime (\beta_c)$ is the derivative of $s$ as a function
of $\beta$. In Fig.5 a best fit by least squares method is reported
yielding $\nu=0.44\pm 0.02$. Here the error is the statistical one
coming out from the linear fit.

Of course this Monte Carlo calculation is far from being the best
numerical simulation which can be achieved by modern computing
machines. Our sample sizes are relatively small and a slight shift of
the critical point cannot be ruled out. However the estimates for
the critical temperature and exponent are accurate enough for a
comparison with experimental findings and for a check of the
analytical results of the previous sections, and that is
just what we needed a the moment. More refined calculations are called
for in order to establish more accurate predictions.

\section{Discussion}

Here we summarize and discuss
the main findings of the previous sections. According to
mean-field and finite-size scaling the three-dimensional
molecular model has a second
order continuous transition from an isotropic disordered high-temperature
phase to an anisotropic two-dimensional ordered low-temperature phase.
The $d=3$ realization of the model is the one which more closely describes
real molecular systems. For this reason the $d=3$ model has been
studied by the variational Migdal-Kadanoff method and by numerical
Monte Carlo simulation. The transition point is characterized by a
diverging correlation length according to the one-parameter scaling
hypothesis which seems to be fulfilled as shown by the data of
the previous section. On the other hand the $d=2$ model is special
by itself for its equivalence to the two-dimensional Ising model,
and for the existence of exact analytical results. Thus the $d=3$ model
can be seen as a non-trivial extension to higher dimension of the
two-dimensional Ising model.  Here ``non-trivial'' means that the
$d=3$ molecular model does not belong to the universality
classes of the standard $d=3$ extensions
of the Ising model (three-dimensional Ising and Potts models).
The difference is evident from a comparison of the ground state $T=0$
configurations: highly degenerate and anisotropic in the molecular
model (with a two-dimensional character even for higher dimensions);
with a small degeneration and fully isotropic in the Potts models
(including the Ising one as a special case). By considering
the two-dimensional
character of the low-temperature phase, the molecular
model could be thought to belong to the universality class of the
simple two-dimensional Ising or three-states Potts models. However in
the high temperature unbroken-symmetry phase
the molecular model is fully isotropic and has a three-dimensional
character.

A formal proof of such statements comes from a comparison of the
critical exponents. For the three-dimensional molecular model the
finite size scaling calculation of the previous section yields
$\nu=0.44$ to be compared to the two-dimensional two-state
(Ising) and three-state Potts models whose exponents are
$\nu=1$ and $\nu=0.83$ respectively\cite{Itz}, to the 
three-dimensional Ising model whose exponent is $\nu=0.64$\cite{Itz},
and to the three-state three-dimensional Potts model which
is known to undergo a first-order transition\cite{wu,janke}.

The molecular model belongs to a new universality class which
is characterized by a sort of dimensional transmutation. In fact
order takes place in chains which are arranged in layers, and the
disorder-order transition requires a decrease of the effective
dimensionality of the system. In the ordered phase the molecules
are correlated inside layers, but there is no correlation between
molecules which belong to different layers. This understanding of
the ordered phase is in agreement with our finding that the
two-dimensional Migdal-Kadanoff variational method for a single
layer yields a better prediction for the critical point than the
three-dimensional method applied to the whole lattice. On the
other hand the very same two-dimensional variational method provides
a convenient analytical tool for describing the generic
$d$-dimensional molecular model by a straightforward generalization.

From
such arguments the critical point has been given an upper bound
by the variational method which yields $\beta_c=0.61$, while a lower
bound is usually provided by mean-field that in this $d=3$ case
gives $\beta_c=0.375$. The numerical estimate of the previous
section $\beta_c=0.53$ fits nicely inside such bounds.

Having discussed some formal aspects of the molecular model and
few approximate methods which throw some light on its phase
transition, we would like to make contact with the phenomenology.
Our main finding - that the order-disorder transition described
by the model belongs to a new universality class - deserves some
experimental test. Transitions of this kind have been observed in
several systems, as discussed in the introduction. Since the critical
properties should not depend on the microscopic details of the system
we expect that the simple molecular model could predict the correct
critical exponent of real orientational transitions occurring in
complex real molecular systems expecially under pressure.
New experiments are called for in
order to test such ideas and explore this broad universality class.

\begin{figure}
\caption{An allowed configuration for $d=2$.}
\end{figure}

\begin{figure}
\caption{Ground state configurations for the two-dimensional
{\it attractive} (a) and {\it repulsive} (b) models.}
\end{figure}

\begin{figure}
\caption{Numerical solutions of the two-dimensional Migdal-Kadanoff
equations for a single layer of the three-dimensional molecular model. 
The critical temperature $\beta$ is reported
as a function of the field strength $h$. The stationary point is at
$h=h_0=-0.2349$ where $\beta=\beta_c=0.6122$. For $h>-0.226$ there is
no physical solution.}
\end{figure}

\begin{figure}
\caption{The fourth moment $s=\langle\gamma^4\rangle$ versus 
the inverse temperature $\beta$ for $N=15,20,25,30$. For $N=30$
only few points around the critical point have been evaluated.
The curves are a linear interpolation between points and are reported
as a guide for the eye.}
\end{figure}

\begin{figure}
\caption{linear fit for the critical exponent according to Eq.(\ref{fit}).
The points have been evaluated for $N=15,20,25,30$.}
\end{figure}

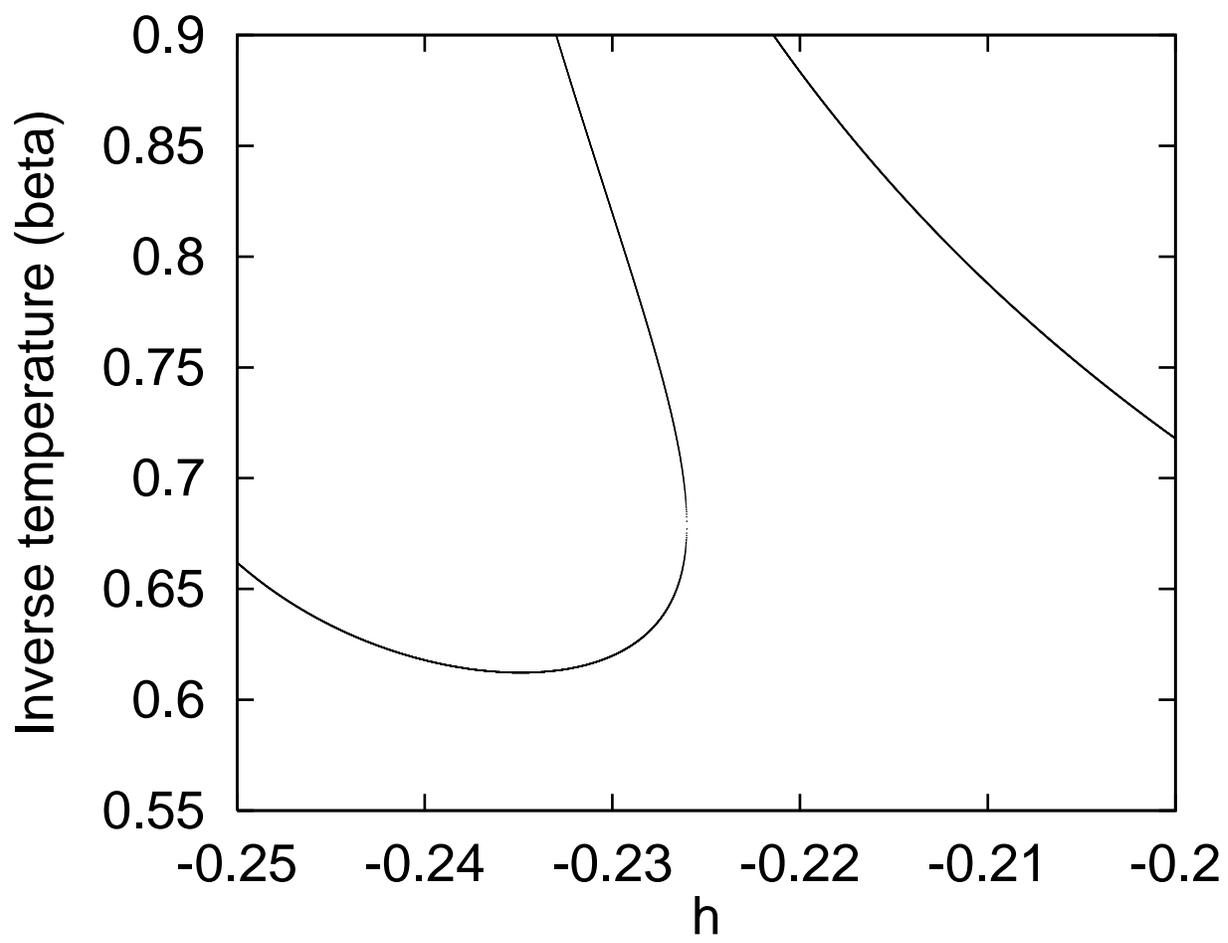
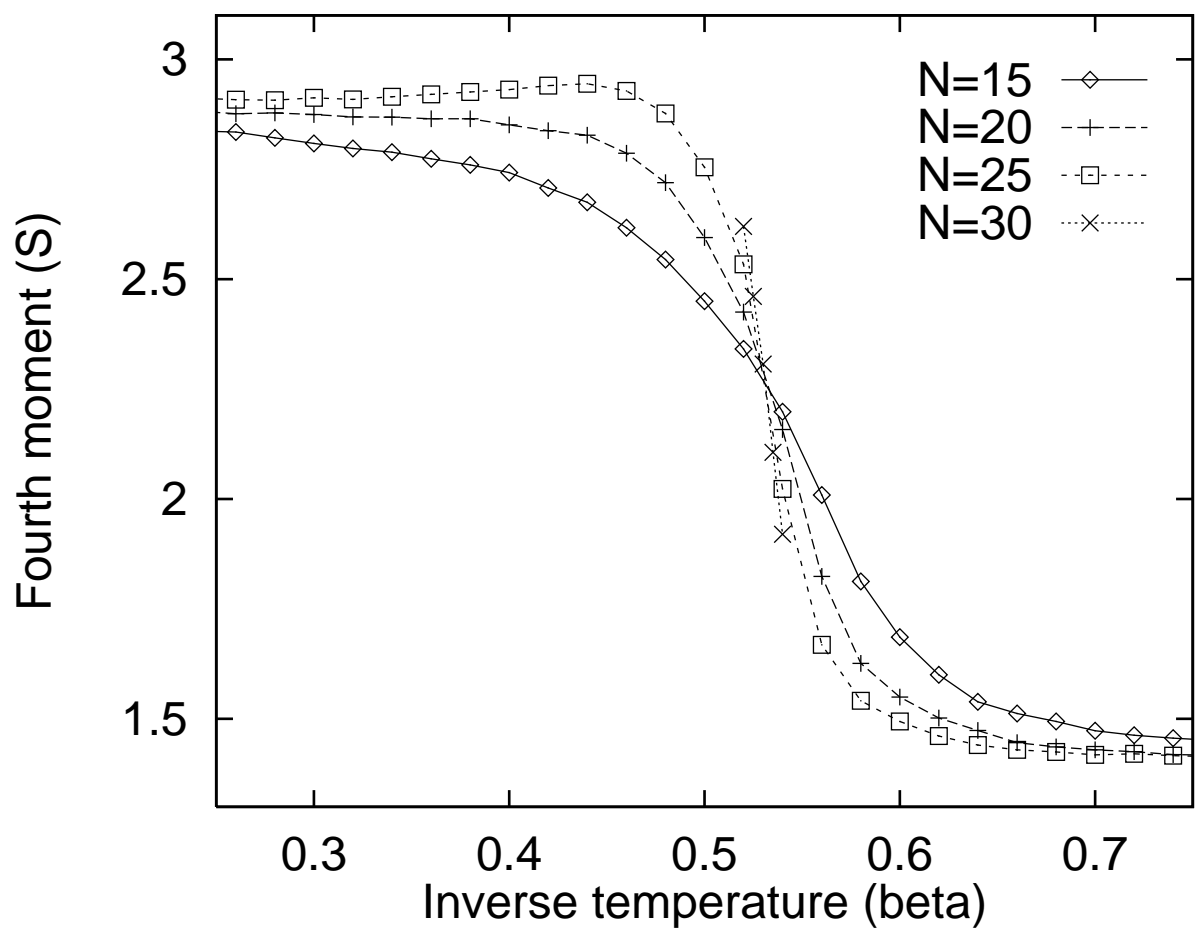
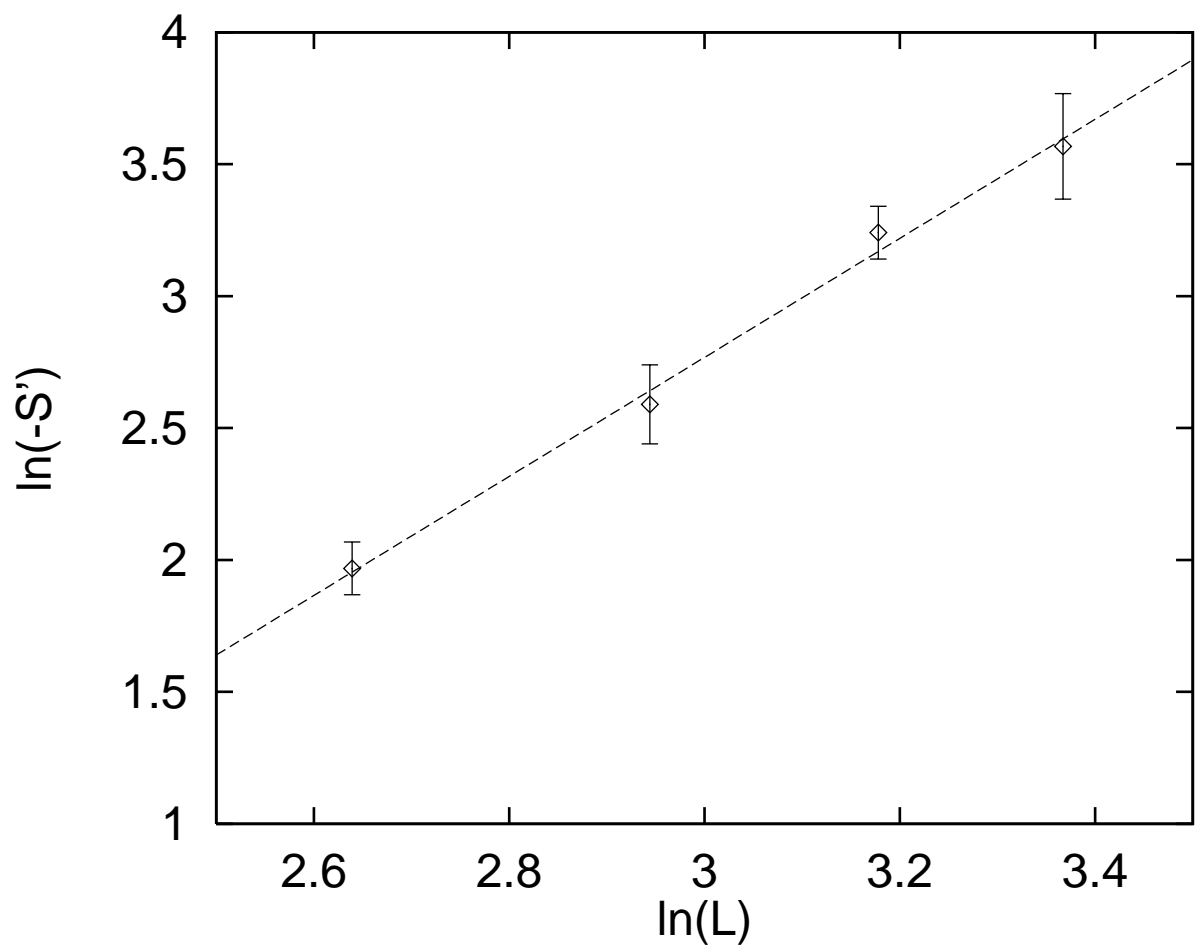
\end{document}